\documentclass[twocolumn,showpacs]{revtex4}
\usepackage{graphicx}
\usepackage{dcolumn}
\usepackage{amsmath}
\begin{document}

\title{A Lattice Model For the Kinetics of Rupture of Fluid Bilayer
Membranes}
%\date{Jan 2003}
\date{\today}

\author{Luc Fournier and B\'ela Jo\'{o}s}
\email{bjoos@uottawa.ca}
\affiliation{Ottawa Carleton Institute of Physics\\
 University of Ottawa Campus\\ Ottawa, Ontario, Canada, K1N-6N5\\}

%\begin{center}
%{Submitted to {\bf Physical Review E.}}
%\end{center}
\begin{abstract}
\vspace*{0.2cm} We have constructed a model for the kinetics of
rupture of membranes under tension, applying physical principles
relevant to lipid bilayers held together by hydrophobic
interactions. The membrane is characterized by the bulk
compressibility (for expansion) $K$, the thickness $2h_{t}$ of the
hydrophobic part of the bilayer, the hydrophobicity $\sigma$ and a
parameter $\gamma$ characterizing the tail rigidity of the lipids.
The model is a lattice model which incorporates strain relaxation,
and considers the nucleation of pores at constant area, constant
temperature, and constant particle number. The particle number is
conserved by allowing multiple occupancy of the sites. An
equilibrium ``phase diagram'' is constructed as a function of
temperature and strain with the total pore surface and
distribution as the order parameters. A first order rupture line
is found with increasing tension, and a continuous increase in
proto-pore concentration with rising temperature till instability.
The model explains current results on saturated and unsaturated PC
lipid bilayers and thicker artificial bilayers made of diblock
copolymers. Pore size distributions are presented for various
values of area expansion and temperature, and the fractal
dimension of the pore edge is evaluated.
\end{abstract}
 \maketitle

\newpage
\section{Introduction}
Fluid bilayer membranes made of lipids separate the contents of
cells from their surroundings. The stability of those membranes
under external perturbations is consequently of vital importance.
In particular, if ruptured the functions of the cell will be
disabled. Naturally, a red cell bilayer membrane may rupture to
free hemoglobin, protein responsible for oxygen transportation in
blood. This particular rupture, called blood cell hemolysis is
achieved through thermal swelling of red cells \cite{DJ84}.
Rupture or lysis can be achieved in a number of ways. Most
experiments involved application of electric fields which produce
compressive forces through the capacitor effect.
\cite{HH79,AA79,NH89,WW93,W99,HM96}. More recently a number of
other ways have been used: mechanically, by suction through a
pipette \cite{OR00}, or extrusion through pores \cite{HF98}; and
by osmotic swelling of the cell \cite{EM93,MC93,TD75}. With the
rapid progress in microscopic manipulation techniques, pore
formation can also serve a positive purpose in drug delivery and
gene therapy \cite{RK95,L95,CS98}. The number of experimental
studies of the mechanical properties of the cell is rapidly
increasing and we only quoted a few examples of recent work
\cite{B01}.

For the fluid bilayers, there is a significant degree of consensus
on the order of magnitude of the quantities involved in membrane
rupture. For phosphatidylcholine (PC) bilayers, area expansion
seems to be only 2{\%} to 4{\%} before rupture. The corresponding
external tension is of the order of 10$^{ - 3}$ to 10$^{ - 2}$ N/m
\cite{NH89,WD85,EN87,OR00,RO00}. It also seems that in liquid
membranes, rupture occurs via the nucleation of holes. Those
observations are consistent with what is expected from structural
considerations. Solid membranes as they go from brittle to
ductile, experience rupture kinetics which evolve from processes
dominated by crack formation to hole formation \cite{ZJ97}. More
precisely, as the surface tension increases, the membrane will
first be weakened by the apparition of small unstable pores with
life-times as short as a few nanoseconds. At a critical tension, a
large stable pore opens, relaxing almost entirely the membrane.
This is rupture. Pore sizes can range from nano- to micro-meter in
radius. In electroporation experiments, pores can last several
microseconds \cite{HH79,AA79,NH89,WW93,AS98}. Using mechanical
means, namely micropipette extrusion on vesicles, pores can be
kept open for several seconds before resealing \cite{ZN93,SM96}.

Several models of membrane rupture have been developed over the
years. Most of them derive from a model suggested by Litster a
quarter of a century ago \cite{L75}. It explains the stability in
terms of a surface energy and a pore edge energy. The model
defines a critical pore size and an energy barrier for the
creation of an irreversibly growing pore. It has been extended and
applied by other groups in particular to electric breakdown
situations \cite{AA79,BW91,FW94}. Shillcock and Boal investigated
the effect of temperature on membrane stability \cite{SB96}.
Temperature increases the entropy of the pore lowering its free
energy. They show that even at zero tension, edge energy is
required for stability. Netz and Schick have also developed a
mean-field theory of the fluid bilayers as stacks of diblock
copolymers \cite{NS96}.

In our work, we derive the energy of a finite size membrane under
lateral expansion. We then consider pore creation as a thermally
activated process \cite{EL00}. There is an important entropic
element implied in a nucleation process. The model is for a
bilayer whose relevant physical quantities have their origin in
the hydrophobicity of the lipid tails \cite{JI85,WE97}. These
quantities are the area compressibility (for expansion) $K$ which
results from the increased exposure of the hydrophobic tails as
the membrane is stretched, the edge tension \textit{$\lambda $},
(or edge energy per unit length) which is the result of the
exposure of these tails to water along the edge of the pore, and
the rigidity $\gamma$ of the tails.

In the model, the bilayer is characterized by an energy per site,
essentially one molecule in size. It incorporates stress
relaxation as a pore is created and grows. Pore-pore interactions
are automatically taken into account. An equilibrium phase diagram
is constructed determining regions of no \textit{pores},
non-critical pores (or \textit{protopores}), and \textit{rupture}.
The phase diagram is in terms of the temperature and the area
expansion of the membrane. The critical temperatures scale with
the strength of the hydrophobic interaction which to first order
is linked to the length of the tails of the lipids. This
hydrophobic interaction is obtained from parameters appropriate to
pure phosphatidylcholine (PC) \cite{OR00,RO00}. Our model applies
to low strain rates since the model assumes that the membrane
remains in mechanical equilibrium as it is stretched. We predict
the rupture scenario for normal membranes, membranes made of
unsaturated lipids, and very thick membranes.

The following two sections develop the model. We first lay its
physical foundations, and then present our Ising like model, which
is solved using Monte Carlo simulations. We continue with the
results and finish with a discussion of the possible extensions of
the model.

\section{Model for membrane rupture}
Upon area increase, stress will build up in the membrane. As
mentioned earlier, being a liquid the most likely stress release
mechanism will be the formation of holes or pores. Let us imagine
that a hole has appeared through thermal induced stress
fluctuations. The question will be whether the gain in energy
occasioned by the relaxation of the surfactants will
counterbalance the energy loss through exposure of the tails of
the molecule to the solvent. If this is the case, the hole will
continue to grow into a large pore that will permit the system to
relax almost entirely. For now, we assume the process to be purely
planar. Whether fluctuations in the third dimension are important
is an open question. At first thought they do not seem
predominant. Assuming an elastic regime for small expansions of
the membrane on a lattice of total relaxed area $a_{m}$ and
molecular area $a_0$, the energy cost associated with the
stretching the membrane is given by
\begin{equation}
\label{eq1} E_m = \frac{1}{2}Ka_m \left( {\frac{\Delta a}{a_0}}
\right)^2,
\end{equation}
where $K$ is the compressibility (for expansion) and $\Delta a = a
- a_0$ the excess area per molecule. For small expansion, the
compressibility can be found experimentally from various
techniques described previously. The apparent compressibility is
the measure of the slope of the stress-strain curve of a membrane
under a small tension, that is the slope of the curve
\begin{equation}
\label{eq2} \tau = K\left( {\frac{\Delta a}{a_0}} \right),
\end{equation}
where \textit{$\tau $} is the surface tension. However, since
those membranes are fluid, thermal transverse undulations are
usually present. For small stretches, a fraction of the tension
does not induce any strain, but rather brings back the membrane in
its plane \cite{ER90}. The real compressibility is obtained by the
slope of the stress-strain curve, but only in the regime where the
undulations have been ironed out. For many lipids, this
compressibility is about 243 mN/m \cite{RO00}.

When a single pore is inserted, some of this stress will be
relieved. The relative change in the area per molecule becomes:
\begin{equation}
\frac{\Delta a}{a_0}={\frac{\Delta a_m}{a_m} - \frac{a_p}{a_m}},
\label{deltaa}
\end{equation}
where $a_{p}$ is the area of the pore and $\Delta a_{m}$ is the
total expansion of the membrane. But the edge of the new pore will
now be exposed to water. As the tails are hydrophobic, this will
result in an increase of the energy proportional to the perimeter
of the pore. For simplicity, we will assume the pore to be
circular. In the presence of a pore, the energy of a stretched
membrane then becomes
\begin{equation}
\label{eq3} E_m (a_p ) = 2\lambda \sqrt {\pi a_m } \left(
{\frac{a_p }{a_m }} \right)^{1 / 2} + \frac{1}{2}Ka_m \left(
\frac{\Delta a}{a_0} \right)^2,
\end{equation}
 The parameter \textit{$\lambda $}
represents the edge tension (the effective edge energy per unit
length). The first term in equation (\ref{eq3}) is the total edge
energy, which represents the loss of energy on the perimeter of
the circular hole. The second term, the surface energy, is the
loss in energy resulting from the relaxation of the membrane
induced by the hole, which augments the density of particles in
the rest of the membrane. For the edge tension, it can be
rewritten as
\begin{equation}
\label{eq4}
\lambda = 2h_t \sigma ,
\end{equation}
where 2$h_{t}$ is the hydrophobic thickness of the bilayer, and
therefore $h_{t}$ is the length of the lipid tails. And $\sigma $,
the hydrophobicity, is the hydrophobic energy of the lipid tails
per unit area. With the usually quoted hydrophobicity of 40 mN/m
\cite{JI85,WE97}, the edge tension $\lambda $ is of the order of
$10^{ - 7}$~mN \cite{ZN93,WE97}. The experimentally measured
values range from $1$ to $4 \times 10^{-8}$ mN \cite{ZN93}.

In the limit of large systems, the energy barrier to rupture is
given by $\lambda ^2\pi / K(\Delta a_m / a_m )^{ - 1}$ and occurs
for a pore area $a_p = (\lambda a_m \sqrt \pi / K)^{2/3}$. When
$a_{p}$ exceeds that critical value, the pore will grow till the
membrane reaches the minimum energy configuration near
$a_{p}=\Delta a _m$. This argument predicts a barrier height of
the order of $10^{4}k_{B}T_{room}$ ($k_{B}T_{room}=4.14 \times
10^{-18}$mJ) for experimentally observed stretches, such as
$\Delta a_m/a_m=4${\%} and a membrane of total area $1\mu
$m$^{2}$, typical for PC membranes. Even if the entropic
components were very significant as mentioned above, it is not
sufficient to decrease significantly this large barrier. Moreover,
according to these arguments, rupture should only happen at
stretches around $\Delta a_m/a_m=50${\%} !

In our opinion, the problem lies in an over-estimation of the line
energy due to two factors. Firstly, when membranes expand, the
tails get exposed to water, so $h_t$ is not a constant (especially
not at $50 \%$) . It decreases with increased stretching of the
membrane. Secondly lipid bilayers are permeable and this
permeability may be affected by the expansion of the membrane. So
the hydrophobicity $\sigma$ used in the computation of the line
energy will be smaller than the ideal value quoted above. We call
$\sigma$ an {\it apparent} hydrophobicity and the $\sigma$ given
above the {\it pure} hydrophobicity $\sigma _{pure}$. $\sigma$
will be an adjustable parameter.

\subsection{Compressibility, hydrophobicity and tail rigidity}
Under expansion, more water will enter the membrane, augmenting
the free energy. The compressibility (upon expansion) $K$ has
therefore its origin in hydrophobicity. Moreover, the rigidity of
the acyl tails controls the amount of water that penetrates the
membrane at a given stretch. There is therefore a very close
relationship between surface energy, compressibility and tail
rigidity. We will closely follow here some of the arguments of
Wortis and Evans \cite{WE97} (themselves based partly on
Israelachvili's book \cite{JI85}) for what pertains to the
compressibility but adding the concept of the rigidity of the
tails.

To estimate $K$,  we must understand the balance of forces that
sets the area of the lipids $a$. The repulsive force between
lipids is strong at short distances, but falls off rapidly as $a$
increases \cite{WE97}. This repulsion is modelled by a potential
of the form $D/a$, where $D$ is a positive constant. On the other
hand, the attractive part of the potential is a direct consequence
of hydrophobicity. More water can enter in contact with the
hydrocarbon chains as the lipids are separated from one another.
In earlier work \cite{JI85,WE97} a linear regime is assumed for
small expansions of the cross section area of a lipid $a$, and the
energy per molecule is written as $\sigma(a-a_{0})$, where $a_{0}$
is the equilibrium area. This assumes that the tails are totally
flexible and that the area exposed to water is equal to the
increase in area of the membrane $(a-a_{0})$ . However, as
illustrated in Fig. 1, the lipid tails more likely will only be
able to partially close up the opening. We introduce a geometrical
factor $\gamma >1$ defined so that $\gamma (a-a_{0})$ represents
the actual surface exposed to water per molecule under stretching,
and hence $\sigma \gamma (a-a_{0})$ the increased energy per
molecule. The factor {$\gamma $} is related to the rigidity of the
tails, and therefore from now on, $\gamma $ will be referred as
the tail rigidity. To have a geometric picture of $\gamma $, lets
imagine that the stretching $\Delta a$ is represented by a circle,
and the effective exposed surface by a cone of surface $S$, as
shown in figure 1. $\gamma$ is then the ratio $S/ \Delta a$. A
rigidity of 1 represents tails that are completely flexible since
the expanded lateral surface {$\Delta a$} is equal to the
increased exposure of the tails to water. In contrast, a high
rigidity represents very stiff chains, such that a small stretch
can lead to a full exposure of the tails to water. With the
geometry shown in Fig. 1, the length $h_e$ of the tails exposed to
water can be written as :
\begin{equation}
\label{eq5} h_e = \gamma r\left( \frac{\Delta a}{a_0} \right)^{1 /
2},
\end{equation}
where $r = \sqrt {a / \pi } $ is the radius of the lipids.  The
non-exposed length $h_{ne}=h_{t}-h_{e}$ should be used in Eq.
(\ref{eq4}) to calculate the line energy when a pore is nucleated
at a certain stretch. Fixing the rigidity $\gamma$ for a certain
type of lipid determines $h_{e}$ and $h_{ne}$ as a function of the
effective stretch.

If we combine both the attractive and the repulsive parts of the
potential, we get a general expression for the energy per
molecular site of the membrane:
\begin{equation}
\label{eq6} U(a) = 2\sigma \gamma (a - a_0 ) + D(a^{ - 1}) + U_0.
\end{equation}
The factor of $2$ in the first term comes from the fact that we
are dealing with a bilayer. The requirement $\left. {dU / da}
\right|_{a_0 } = 0$ leads to the relationship $D=2\sigma \gamma
a_{0}^{2}$. It must be emphasized that \textit{$\sigma $} here
represents the interaction of the lipids when present in the
membrane, and therefore refers to the \textit{apparent}
hydrophobicity, in opposition to the pure hydrophobicity $\sigma
_{pure}$ discussed earlier in this section. By comparing the
curvature of this potential around the minimum with the surface
energy of Eq. (\ref{eq1}), we obtain
\begin{equation}
\label{eq7}
K = 4\sigma \gamma .
\end{equation}
Thus, there is a close relationship between the compressibility
$K$, the hydrophobicity $\sigma$ and the rigidity $\gamma$. In our
simulations, we use the experimentally known values of
compressibility $K$, and set the rigidity $\gamma $ to obtain
rupture near 4{\%} for the PC type bilayers. With the use of Eq.
(\ref{eq7}), we shall demonstrate in section \ref{res-rig} that
the apparent hydrophobicity $\sigma$ is considerably smaller than
the pure one $\sigma_{pure}$, which ignores the water already
present in the membrane. With $\sigma_{pure}$ the line energy is
over-estimated as we saw above.

\section{The calculation}
What we are trying to simulate is a thermally activated nucleation
process, it is therefore natural to consider a model amenable to a
Monte Carlo simulation. For ease of computation we consider a
lattice model similar to the well-known Ising model for binary
mixtures, and consider the equilibrium phase diagram of the system
as a function of temperature and area expansion. This will reveal
the expected scenarios of rupture as the membrane is expanded
slowly, quasi-statically, so that the membrane remains in
thermodynamic equilibrium at every step of the way. We will limit
ourselves to two-dimensions, and to represent an isotropic liquid,
we will use a hexagonal lattice (6 nearest neighbors). We begin by
presenting the standard binary mixture model (SBMM) as applied to
our problem and then describe the modifications and
simplifications that we have made to it to incorporate the stress
relaxation that occurs when a hole is created. In the SBMM, every
site can be in either of two states, in our case ``a'' or ``h'',
representing respectively sites occupied by a phospholipid or a
hole. The site occupancy variable $s_{i}$, is equal to 1 if the
site is an ``a'' site and 0 if it is a ``h'' site. The indices $i$
and $j$ run from 1 to $N$, where $N$ is the total number of sites.
Let $J_{aa}(r_{ij})$, or its compact form $J_{ij}^{aa} $, be the
interaction between two lipids separated by a distance
$r_{ij}=|{\bf R}_{i}-{\bf R}_{j}|$, where ${\bf R}_{i}$ is the
position vector for site $i$. In a similar way, $J^{hh}(r_{ij})$
and $J^{ah}(r_{ij})$ are defined. The microscopic Hamiltonian $H$
consists of the sum of all interactions:
\begin{eqnarray}
\label{H} H = \frac{1}{2}&\sum\limits_{ij}& \left[ J_{ij}^{aa} s_i
s_j + J_{ij}^{hh} (1 - s_i )(1 - s_j )\right. \nonumber \\ && +
\left. J_{ij}^{ah} \left[ {s_i (1 - s_j ) + s_j (1 - s_i )}
\right] \right] \label{eq8} .
\end{eqnarray}
In a conventional canonical binary mixture, one would expand Eq.
(\ref{eq8}) and eliminate the constant terms and the single sums
since the number of each type of particle is conserved. In our
binary mixture model, we keep the number $N$ of particles (the $a$
sites) and the total area constant. The number of holes (the $h$
sites), however, can vary. Our model is neither the conventional
canonical or grand canonical version of the binary mixture model.
In the canonical ensemble, occupied ($a$) and empty ($h$) sites
are interchanged to preserve the total number of particles. The
conventional canonical model would be used if we were only
interested in the phase separation of a system, which is not our
case. In the grand canonical ensemble, the only dynamics that
would be observed is the distribution of the holes. Neither model
is suitable to incorporate the relaxation created by the
appearance of a hole.

We start with the SBMM in the canonical ensemble with an initial
total number of particles equal to the number of sites $N$. We now
modify the model to impose the following constraints: the total
area of the membrane stays constant as holes are created and the
total number of lipids --$a$ particles- stays constant. These lead
to the following change in the dynamics. Occupied sites, when
holes are created, now contain more than one particle, actually
$N/(N-n_{h})$, where $n_{h}$ is the number of hole sites at a
given time. The area of the occupied and hole sites stay constant.
The nucleation of holes therefore relaxes the membrane and reduces
the interparticle distance $r_{ij}$. This relaxation represents
the surface energy and is the $J_{ij}^{aa}$ term in Eq.
(\ref{eq8}).

On the other hand, the interaction energy between holes and
particles $J_{ij}^{ah} $ represents the line energy of pores. The
occupancy of sites being larger than one in such a case, there are
more particles on the edge than the actual number of sites. As we
will discuss in section \ref{line-energy}, this will lead to a
correction in the line energy. Finally the hole-hole interaction
$J_{ij}^{hh}$ is set to zero.

\subsection{The surface energy \label{surf-energy}}
The surface energy is not calculated using the first term in Eq.
(\ref{eq8}) but directly using the occupied area in the membrane.
With $n_{h}$ hole sites, the interparticle distances have to be
rescaled to the new values
\begin{equation}
\label{eq9}
\tilde {r}_{ij} = \left( {\frac{N - n_h }{N}} \right)^{1 / 2}r_{ij} ,
\end{equation}
where the $r_{ij}=\vert {\bf R}_{i}-{\bf R}_{j} \vert $ are the
initial intermolecular distances. As discussed earlier, the
nucleation of holes changes the density of lipids in the membrane
but conserves their number. Since we assume uniform relaxation at
each step of the simulation, the relative change in area per
lipid, defined in Eq. (\ref{deltaa}), can be re-written as
\begin{equation}
\frac{\Delta a}{a_0}={\frac{\Delta a_m}{a_m} -
\frac{a_p}{a_m}}={\frac{\Delta a_m}{a_m} - \frac{n_h}{N}},
\label{deltaam}
\end{equation}
where $a_p$ is now the total pore area, and not just the area of
one pore.  The surface energy can be calculated directly using the
compressibility from the relation,
\begin{equation}
\label{eq10} E_s = \frac{1}{2}Ka_m \left( \frac{\Delta a}{a_0}
\right)^2 .
\end{equation}

It is clear from this last equation that if the relative number of
holes $n_{h}/N$ completely relaxes the imposed stretch on the
membrane $\Delta a_m/a_m$, the surface energy vanishes, as
expected.

\subsection{The line energy \label{line-energy}}

As previously mentioned, the edge energy represents the
hydrophobic contacts along the perimeter of a pore, measured by
the hole-lipid interaction $J_{ij}^{ah} $. The edge energy
therefore depends on the location of the holes in the lattice. The
sum of those interactions cannot be reduced to a simple form like
the surface energy.

As we are working on a hexagonal lattice, $J_{ij}^{ah} $ consists
of the energy of exposing one sixth of the hydrophobic surface of
a lipid to water, multiplied by two for the bilayer. It is a first
neighbor interaction. $J_{ij}^{ah} $ can be written as
\begin{equation}
\label{eq11} J_{ij}^{ah} = J^{ah}=\left\{ {{\begin{array}{cl}
 {\frac{2}{3}h_{ne} \sigma \sqrt {\pi a_0} } & \ \ \ \mbox{if}\,r_{ij}
\,\mbox{is}\,\mbox{1}^{\mbox{st}} \,\mbox{neighbor}\\
 0 & \ \ \ \ {\,\,\mbox{otherwise}}
\end{array} }\,\,\,} \right.
\end{equation}
where $a_0$ is the cross section of the lipids, and
$h_{ne}=h_{t}-h_{e}$ the hydrophobic length of the lipid tails
that are not exposed to water ($h_t$ and $h_e$ are defined in Eqs.
(\ref{eq4}) and (\ref{eq5}) respectively) . The hydrophobic energy
of the exposed height $h_{e}$ is the origin of the surface energy.

In the simulation the occupancy variables $s_{i}$ remain equal to
either $0$ or $1$. So the edge energy as calculated by the program
is
\begin{equation}
\label{eq13} E_{edge}^{\mbox{\scriptsize model}} = J^{ah}L^{ah},
\end{equation}
where $L^{ah}$ is the number of lipid-hole interactions (between
sites).

However, every time a hole site is created, the density of the
particles and hence the interparticle distance changes as
expressed in Eq. (\ref{eq9}), in particular the nearest neighbor
interparticle distance $r$ . There are therefore more particles on
the edge of a pore than the actual number of sites. The edge
energy per unit length can be written as $J^{ah} / \tilde {r}$,
where $\tilde {r}$ is the scaled nearest neighbor interparticle
distance. And hence the correct expression for the line energy is:
\begin{equation}
\label{eq12} E_{edge} = \frac{J^{ah}}{\tilde {r}}L^{ah}r.
\end{equation}
Using Eq. (\ref{eq9}), Eq. (\ref{eq12}) can be written as
\begin{equation}
\label{eq14} E_{edge} = \left( {\frac{N}{N - n_h }} \right)^{1 /
2}E_{edge}^{\mbox{\scriptsize model}},
\end{equation}
or in an expanded form,
\begin{equation}
\label{eq15} E_{edge} = {1 \over 2} \left( {\frac{N}{N - n_h }}
\right)^{1 / 2}\sum {J_{ij}^{ah} \left[ {s_i (1 - s_j ) + s_j (1 -
s_i )} \right]} .
\end{equation}

\subsection{Programming and Analysis}
To create and recover holes in the membrane up to an equilibrium
point, we use the Metropolis algorithm in the Monte Carlo
simulation. The mass of a hole is zero, but the mass of the system
or number of particles is conserved through the renormalization of
Eq. (\ref{eq5}). Although the occupation of sites changes as the
simulation evolves, detailed balance is satisfied because each
configuration of pores has a uniquely defined energy. This energy
does not depend upon the path followed to reach it. The transition
probability between a given initial and final configuration will
also be unique and reversible. If we choose an occupied site with
the same probability than an empty site, that is, we select
particles or holes according to their surface distribution; the
acceptance rule will be given by
\begin{equation}
\label{eq16} \mbox{acc}(i \to f) = \min \left(1,\exp \left[ { -
\left( {\tilde {H}_f - \tilde {H}_i } \right) / k_B T}
\right]\right),
\end{equation}
in order to respect detailed balance. This acceptance rule is the
usual one for a Monte Carlo simulation in the canonical ensemble.
Finally, periodic boundary conditions are considered on the
hexagonal lattice.

The approach is highly efficient and large systems can be studied,
up to 10$^{7}$ particles. However, systems of 10$^{4}$ particles,
which correspond to small vesicles, are normally sufficient.
Usually within 1000 samplings per site, the equilibrium state is
easily reached.

\section{Results}
The types of membranes we study can be grouped in three main
categories: saturated lipid bilayers (typical), unsaturated lipid
bilayers, and long chain diblock copolymer bilayers. Membranes
with unsaturated lipids may be found in biological systems. The
long tail diblock copolymers, however, were synthesized by
Bermudez et al., and then used to built vesicle shape polymersomes
\cite{Bermudez}. In Table 1 are grouped the important parameters
for our study. Those values are the total number of unsaturated
bonds, the compressibility $K$, the membrane core thickness
$2h_{t}$, the permeability, the critical tension for rupture $\tau
^*$, and the corresponding stretching $\Delta a_m$*$/a_m$. Values
for lipid membranes were taken from Olbrich et al. \cite{OR00} and
Rawicz et al. \cite{RO00}. The unsaturated lipid membranes differ
from the typical membranes in the stiffness of their tails while
the diblock copolymers simply have longer tails.

\begin{table}[htbp]
\caption{Compressibility and other known physical parameters of
the different types of membranes studied.   These parameters are:
the total number of cis-unsaturated bonds in the lipid tails
(''{\it uns}.''), the compressibility $K$ , the thickness of the
tails $2 h_t$, the permeability $P$, the critical
 tension for rupture $\tau *$, and corresponding area stretching
 $\Delta a_m */ a_m$.
 Values for typical (i.e. with none or only one unsaturated bond) and
unsaturated lipids
 are from Olbrich et al. \cite{OR00} and Rawicz et al. \cite{RO00},
 whereas the values
 for long tail diblock copolymers are taken from Bermudez et al.
 \cite{Bermudez}.}
\begin{tabular}{llllllll}
\hline & $uns.$&  $K$ &
2$h_{t}$  & $P$  & $\tau *$ & $\Delta a_m*/a_m$   \\
\multicolumn{2}{l}{membrane} & (mN/m)&(nm)& ($\mu
$m/s)&(mN/m)&({\%})\\ \hline
 \multicolumn{3}{l}
 {Typical (average)} & & \\&
 0, 1& 243$\pm $24& 3,0$\pm $0,1& 35$\pm $7& 9$\pm $2&
3,9$\pm $0,9 $^{1}$ \\ \multicolumn{3}{l}
 {Unsaturated lipids} & & \\
  SLPC:0/2& 2& 243$\pm $24& 3,0$\pm $0,1& 49$\pm
$6& 4,9$\pm $1,6&
2,0$\pm $0,5 $^{1}$ \\
  DLPC:2/2& 4& 243$\pm $24& 3,0$\pm $0,1& 91$\pm $24&
5,1$\pm $1,0&
2,1$\pm $0,4 $^{1}$ \\
  DLPC:3/3 & 6& 243$\pm $24& 3,0$\pm $0,1& 146$\pm $24&
3,1$\pm $1,0&
1,2$\pm $0,4 $^{1}$ \\
\multicolumn{3}{l}
 {Long tail copolymers} & & \\
  OE7& 0& 102$\pm $10& 8$\pm $1&
---&
21$\pm $3 $^{1}$&
21$\pm $2 \\
  OB9& 0& 102$\pm $10& 11$\pm $1&
---&
28$\pm $4 $^{1}$&
28$\pm $3 \\
  OB18& 0& 102$\pm $10& 21$\pm $1&
---&
40$\pm $5 $^{1}$&
40$\pm $4 \\
\hline
\end{tabular}
Abbreviations: SLPC:0/2 = 1-stearoyl-2-linoleoyl$_{ cis at 9, 12
}$-phosphatidylcholine; \\DLPC:2/2 = dilinoleoyl$_{ both cis at 9,
12 }$-phosphatidylcholine;\\ DLPC:3/3 = dilinoleoyl$_{ both cis at
9, 12, 15 }$-phosphatidylcholine;\\ OE7 =
ethylenoxide$_{40}$-ethylethylene$_{37};$\\ OB9=
ethylenoxide$_{50}$-butadiene$_{55};$ \\ OB18=
ethylenoxide$_{80}$-butadiene$_{125}$. \\ $^{\ref{eq1}}$ Values
obtained using Eq. (\ref{eq2}), and strains or stresses to rupture
from quoted articles. \label{tab1}
\end{table}

In summary (details appear in the following subsection) our
strategy was to first use the above parameters to determine
$\gamma$ the tail rigidity by looking at the effect it has on the
critical rupture tension. $K$ being known, every choice of
$\gamma$ sets also the value of the apparent hydrophobicity
$\sigma$ through Eq. (\ref{eq7}). Experimentally $K$ seems
independent of tail structure for most lipids \cite{RO00}, but the
rupture kinetics are not and this, we assume, is due mainly to
changes in rigidity or $\gamma$. The compressibility of long tail
copolymers is different from the usual $243$ mN/m mainly because
the head group is different, and not because of the tail length
\cite{Bermudez}. We will focus our attention mainly on the typical
bilayers.

Once $\gamma$ and hence $\sigma$ are set, we study the scenario of
rupture through an ``equilibrium'' phase diagram that looks at the
different qualitative changes in the distribution of pores as a
function of stretching for temperatures around $k_{B}T_{room}$.
This corresponds to slow (quasistatic) stretching rates that keep
the membrane in a state of equilibrium with respect to the applied
stress. In the case of a higher rate of expansion, larger
stretches and therefore tensions will be needed to break the
membrane \cite{EL00,EH03}. Because we are considering thermally
activated processes, higher temperatures in this model are
equivalent to weaker interactions. To study the reversibility of
membrane lysis, hysteresis curves were constructed to compare the
number of holes and their distribution as we first stretch the
membrane to rupture, and then let it relax to zero tension. This
study revealed that the rupture transition is first order at room
temperature, with a strong hysteresis. The fractal dimensions of
stable pores were computed at different temperature as a measure
of their entropy and shape. As a prelude we look at system size
dependence.

\subsection{System size dependence
\label{sys-size}} To simplify the discussion, we assume that the
correct tail rigidity has been obtained. In reality, the numerical
value of the tail rigidity depends somewhat on system size.

Bilayer membranes can form vesicles of a variety of shapes. In
nature, they are normally present as spherical vesicles of radii
between $1$ to $10 \mu $m [11-13]. For a given relative stretch
$\Delta a_m/a_m$ the surface energy per particle does not depend
on the size of the lattice. The line energy scales as the length
of pore edge or the sum of the square root of the surfaces of the
individual pores. System size may affect the kinetics because of
the changes in pore size distribution and the fact that pore size
distribution does not scale with system size.

As shown in figure 2, stretching to rupture almost stabilizes to
about $3.9{\%}$ for systems larger than $10^{4}$ particles. For
smaller systems, we observe a significant increase. As expected,
this plateau is mainly due to the increase in the number of pores
at rupture, which raises the line energy (see figure 3).
Furthermore, entropy increases for larger systems, which favors
earlier rupture. These plots are averages with a standard
deviation of 10 runs. It is an interesting and open question to
know if multiple pore rupture is observed in biological membranes.
It must be noted that the number of sites in our simulations
corresponds to half the number of particles, since we are dealing
with bilayer membranes. For what follows, we shall work on a
lattice of 30301 sites (60602 lipids). This corresponds to a total
membrane area of $18 \mu $m$^{2}$ ($a_0\approx 0.6$nm$^2$ for
phospholipids). For a spherical vesicle, it corresponds to a
radius of $1,5 \mu $m, a small natural biological cell.

\subsection{The rigidity and hydrophobicity of lipid tails
\label{res-rig}}
Depending on the rigidity of the tails of a
lipid, more or less water may penetrate the membrane under
stretching. For unsaturated lipids, one would expect higher
rigidities, since the tails are stiffer. Furthermore, if the tails
are rigid, the membrane is also more permeable \cite{RO00}. In
this case, the apparent hydrophobicity of the membrane is lower,
since the difference in energy for the lipid inside and outside
the membrane is smaller. For small variations of these parameters,
we can see from Eq. 7 that a saturated and an unsaturated membrane
made out of the same lipid group should have comparable
compressibilities. This was verified experimentally by Rawicz
\textit{et al}. \cite{RO00} in their investigation of the effects
of lipid unsaturation on membrane elasticity. This group has found
that unsaturation has no effect on compressibility. Nevertheless,
unsaturation is still an important issue for membrane rupture, as
unsaturated lipids are less flexible than saturated ones. Olbrich
et al. \cite{OR00} have discovered that rupture tension was only
affected for bilayers made with lipids of two or more alternating
cis-double bonds (C=C-C=C) in one or both chains. The three
degrees of unsaturation that we study are as shown in Table 1:
lipids with one tail having two alternating cis-double bonds (1
pair), both tails having two alternating cis-double bonds (2
pairs) and both tails with three alternating cis-double bonds (4
pairs). All these unsaturated lipids also have 18 carbon atoms in
each tail. For the long polymers, the line energy is directly
proportional to the length of the lipid tails. It is therefore
natural to expect higher tensions at rupture for longer chains.
Membrane rupture of diblock copolymers membranes will also be
studied.

\begin{table}[htbp]
\caption{Tail rigidity and apparent hydrophobicity of the
different lipids studied. The estimated value for pure
hydrophobicity is also given as a reference.}
\begin{tabular}{llllll}
 \hline
\text{Membrane}& \text{Name}& \text{Uns.}& $\gamma $& $\sigma $ &
$\sigma _{pure}$ \\
&&&& \text{(mN/m)}& \text{(mN/m)} \\ & & & & &\\ Typical&
(average)& 0, 1& 8.0$\pm $1.0& 7.6& 40 \\& & & & &\\ Unsaturated&
SLPC:0/2& 2& 13.0$\pm $3.0& 4.7&
$ \approx 40$ \\
 lipids& DLPC:2/2& 4& 13.0$\pm $3.0& 4.7&
$ \approx 40$ \\
 & DLPC:3/3 & 6& 17.5$\pm $0.5& 3.8&
$ \approx 40$ \\
 & & & & &
 \\
 Long tails& OE7& 0& 7.3$\pm $1.0& 3.4&
--- \\
 copolymers& OB9& 0& 8.6$\pm $1.0& 2.9&
--- \\
 & OB18& 0& 13.0$\pm $1.0& 2.0&
--- \\
\hline
\end{tabular}
\label{tab2}
\end{table}

To obtain $\gamma $ and therefore $\sigma $,  of the different
membranes, we fit the tail rigidity to have the corresponding
stretching to rupture. For typical lipid bilayer membranes, we
need a rigidity of $\gamma =8 \pm 1$ to have rupture at $3.9${\%}
(see figure 4). From Eq. (7), this corresponds to a hydrophobicity
of $7.6$.

The plateau near $2{\%}$ stretching in figure 4 for large $\gamma
$'s may seem abnormal at first sight. Even more, just before zero
tension rupture, there is even a slight increase in the stretching
to rupture. This effect is related to the criterion that defines
rupture. Zero tension rupture means that the lipids are so stiff
and the membrane so permeable that stability is not even assured.
Just before this critical rigidity, the membrane is still not very
stable, since the line energy per lipid rapidly decreases with
stretching. This weak interaction between lipids raises the
entropy of the system. For high rigidities, this gain in entropy
will make the rupture scenario more comparable to a melting
phenomenon than a pore creation relaxation. Since we have lost the
ideal scenario of rupture which favors relaxation as opposed to
the creation of line energy, rupture in terms of a few stable
pores is more difficult to achieve. The definition of rupture will
be discussed in more detail in the next subsection.

Nevertheless, this plateau seems consistent with experimental
observations. Unsaturated lipids with 2 and 4 total unsaturated
bonds both have a stretching to rupture near 2{\%} (see figure 4).
This therefore means that even though they have the same
stretching to rupture does not mean they have the same tail
rigidity. In fact, they should not.

For typical membranes, the apparent hydrophobicity $\sigma$ is
about 5 times smaller than the approximated value for $\sigma
_{pure}$ (see table 2). This leads to a line energy of $2.4 \times
10^{-8}$ mN, well with the experimentally measured range of $1$ to
$4 \times 10^{-8}$ mN. For highly unsaturated lipids, this ratio
can be as high as 10 if we suppose that the pure hydrophobicity
has not changed significantly. It would be wrong, however, to say
that 80{\%} of the membrane is flooded. This apparent
hydrophobicity may also take into account other repulsive
energies, such as the head-head repulsion. Moreover, $\sigma
_{pure}$ was approximated from geometrical simplifications, and
might be smaller than 40 mN/m. Nevertheless, this noticeable
difference between the pure and the apparent hydrophobicity
indicates that membrane permeability is not a negligible effect.

Graphics similar to figure 4 were used to obtain the tail rigidity
of the diblock copolymers studied by Bermudez et al.
\cite{Bermudez}. The tail rigidity of OE7 and OB8 is comparable to
the rigidity of a typical lipid tail. However, for OB18, which has
a much longer tail (see Table I) $\gamma $ is considerably higher.
It may be that for very long polymers tail entanglements make the
tails appear more rigid \cite{Bermudez,DW99}.

In the following subsections, we focus on {\it typical} biological
membranes. So $\gamma $ is set equal to $8$.

\subsection{Phase diagram}
A phase diagram was constructed to study the behavior of the
membrane under variations of the temperature and the surface area
(i.e. under stretching). We show the phase diagram in Fig. 5 for a
''typical'' membrane (see table I). The most striking feature is a
''first order transition'' line to a ruptured state. When at a
given temperature, the membrane is stretched, one observes in the
neighborhood of a specific stretch an abrupt increase in the
relaxation which we call {\it rupture}. This occurs through the
formation of a few massive stable pores. The line shown in Fig. 5
corresponds to the point of abrupt relaxation as the tension is
increased. Below this line the nature of the membrane evolves
gradually with temperature $T$. At low $T$, including around
$T_{room}$, there is what we call a {\it stable} state. The
relaxation in this regime is small, with occasionally a few
protopores (see Section \ref{sec-por}). As the temperature is
raised the number and the size of the protopores keep increasing.
The pores become numerous, but are still short lived, and of size
no larger than 10 particles or so.

At temperatures below and near $T_{room}$, rupture is very sharp.
We go directly from a state of high tension where less than $20\%$
of the membrane is relaxed to an almost fully relaxed system
($80{\%}$ and more). At higher temperatures, there are a larger
number of pores present in the membrane at low stretches, but they
do not aggregate till the tension reaches the rupture tension.

The rupture line first decreases with temperature till $1.6
T_{room}$ then rises again. Entropy appears to favor a high
density of dispersed pores. It takes then an increased tension to
force the aggregation of the pores and trigger the full relaxation
of the membrane. This leads to the rise in the rupture tension.
Finally the rupture line ends near $T=2.8T_{room}$, but it is no
longer a first order transition line. At that temperature the
membrane loses stability: {\it stable} pores are created without
any tension applied. The reduced line energy $\lambda /k_B T$
falls below the critical value required to assure membrane
stability. Shillcock and Boal \cite{SB96} vary $\lambda$ in their
tethered beads model. They find a critical value of $\lambda ^* b
/k_B T =1.66$ at constant zero-tension, where $b$ is the average
vertex separation in their model (At $T_{room}$, this means a
critical line energy of about $0.9 \times 10^{-11}$ J/m). We work
at constant area and, in our model, we calculate hole-lipid
interactions and the total line energy is given by $J^{ah}L^{ah}$
(see Eq. (\ref{eq12})) , where $J_{ij}^{ah}$ is the hole-lipid
interaction defined in Eq. (\ref{eq11}). The effective line
tension depends on pore distribution.  With our thermally
activated model the behavior observed near $2.8 T_{room}$ can be
viewed as behavior near $T_{room}$ with a line energy reduced by
the factor $2.8$. Using the distribution of pores in Fig.
\ref{fig-por}c, one can find an effective line energy per site of
approximately $1.1J^{ah}$ (small pores dominate the distribution)
\cite{note}, or a critical line energy per unit length of about
$0.94 \times 10^{-11}$ J/m. Interestingly our lattice model and
Shillcock and Boal's bond flipping model have consistent views on
stability. Mechanically the membrane may fail earlier so we label
the region around $T=2.8 T_{room}$ as {\it unstable}.

At room temperature, the presence of a regime offering $2{\%}$ to
$20{\%}$ unstable pore surface, or small non-stable pores, called
protopores, have experimentally been observed at room temperature
\cite{NH89}. In the next three subsections we look at different
properties of the membrane under tension that give a clearer
picture of the rupture kinetics.

\subsection{Hysteresis \label{sec-hys}}
To show that indeed rupture is ''first-order'', samples were put
through a full cycle of expansion and compression. This is done
again in a quasi-static process, i.e. at each step the system is
allowed to relax. We follow, as the strain was varied, the
variations in the stress (Fig. 6a), the relative pore area (a
measure of the relaxation)(Fig. 6b), and the total number of pores
(a measure of pore interaction and coalescence)(Fig. 6c).

Hysteresis effects are found to be very strong at low temperature.
Rupture and healing occur through nucleation of one pore. At high
temperature, hysteresis is reduced and there are many more pores
present, below and above rupture. At $T_{room}$, the case
illustrated in Figs. 6,  behavior is similar to the high
temperature situation below rupture, but comparable to the low
temperature case above rupture. A considerable number of pores are
present in the membrane at first, but after rupture, only a small
number of large non-fluctuating pores remain.  This is clearly
observable in Fig. 6c as the rapid rise in the number of pores
which precedes the collapse at rupture. The tension at collapse
defines the $T_{room}$ point on the rupture line.

\subsection{Pore size distribution \label{sec-por}}
To further illustrate the difference in pore distribution below
and above rupture, we compare the distribution of pore sizes for
membranes with 2{\%} and 6{\%} strain. We do this at the three
temperatures $T_{room}$, $2T_{room}$, and $3T_{room}$ to
illustrate the change in structure of the membrane with increasing
temperature (see Figs. \ref{fig-por} (a),(b), and (c)). The
distributions are typical snap-shots of the membrane as the Monte
Carlo simulation evolves. For the stable membranes shown, we see
in figures 7a and 7b, the sharp drop in the number of small pores
above rupture. A large pore creates less line energy than a
protopore for the same gain in surface energy, so is energetically
favored. For unstable membranes, the appearance of larger pores in
the system does not reduce the number of smaller ones (figure 7c).
The small pores, present at zero tension at $3T_{room}$ are stable
due to the contribution of entropy to the free energy. This
illustrates from another perspective the instability of the
membrane at 3$T_{room}$.

\subsection{Pore shape \label{sec-sha}}
At different temperatures for a stretching of 3{\%} (just before
rupture), we insert into the membrane a hole at the origin. After
the pore is fully opened and stable, we compute its fractal
dimension $d_f$ from the scaling relationship between the pore
area $A$ and its perimeter $l$ ($A=A(0)l^{d_f}$ where $A_{0}$ is a
proportionality factor).

From the results plotted in figure 8, we clearly see that $d_f$
decreases as the temperature is raised. This is caused by
increased irregularities along the edge of the pore. At
 $T$ = 2.8 $T_{room}$, where rupture occurs spontaneously at zero
tension, one would expect the scaling relationship applicable to a
self-avoiding ring \cite{SB96}. In such geometries, the area of
the pore should scale as $A =A_{0 }l^{ 3 / 2}$. For this case, our
model gives $d_f= 1,54 \pm 0,10$ in agreement with this argument.

\section{Conclusion}
Bilayer membranes are an intrinsic part of all living species.
With the rapid development of biotechnology, membranes can now be
artificially created, and thereafter even modified to form
vesicles of controlled size and properties. Rupture kinetics play
an integral part in the modifications of those vesicles and in
their applications such as in drug delivery.

In spite of their complex molecular structure, lipid bilayer
membranes offer an ideal model system, as hydrophobicity is
responsible for both the edge line energy and the bulk
compressibility (for extension). We have developed the first model
for the rupture of a membrane held together by hydrophobic forces,
which includes the nucleation and growth of pores. Our minimal
model can be used to reproduce current results on saturated and
unsaturated PC lipid bilayers and thicker artificial bilayers made
of diblock copolymers. We have introduced a new quantity called
the rigidity of the tails $\gamma$ which permits the study the
effect of saturation of the lipid tails on the structural
properties of the bilayer. Interestingly an increase in rigidity
produces a decrease in apparent hydrophobicity since $K$ is nearly
constant for a given class of lipids. This may mean an increase in
permeability. The model requires little computer time allowing the
handling of real size vesicles. Entropic and nucleation effects
are included naturally.

The structural integrity of a lipid bilayer is considerably
affected by its composition; biological membrane structure can be
fairly complex. Different types of lipids, cholesterols, and
proteins play an important role in determining membrane stability.
The model has the potential of handling these inclusions.

\section*{Acknowledgements}
The authors wish to thank Michael Wortis, James Polson, David
Boal, Dennis Discher, and Martin Zuckermann for many stimulating
discussions. The work has been funded by the Natural Sciences and
Engineering Research Council (Canada).

\newpage
\begin{figure}
\resizebox{3.00in}{1.88in}{\includegraphics{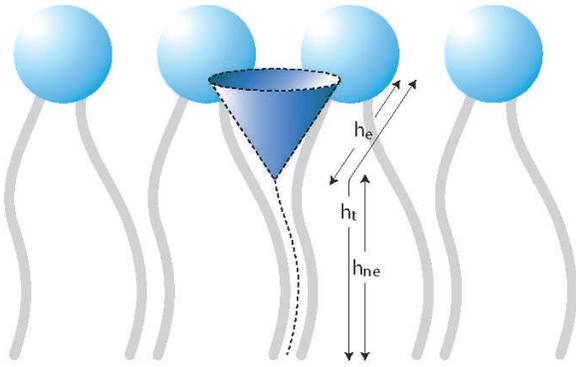}}
%\resizebox{3.00in}{1.88in}{\includegraphics{fjfig1}}
 \caption{
Schematic representation of the total height of the lipid tails
exposed to water at a given expansion. The rigidity $\gamma$ is
the ratio of surface of the sides of the cones to the base. The
total surface exposed to water is expected to be larger than the
surface expansion. }
\end{figure}
\begin{figure}
\resizebox{3.25in}{2.62in}{\includegraphics{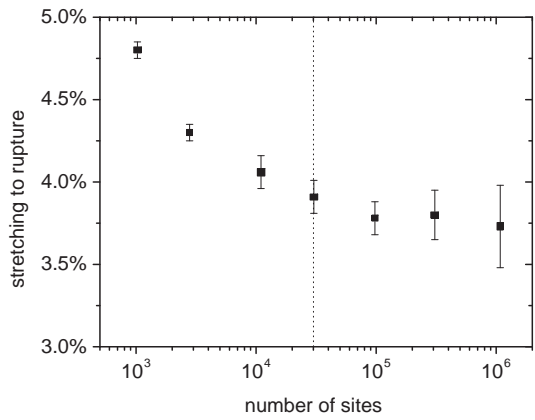}}
\caption{The dependence of the degree of stretching  on the size
of the membrane for rupture at room temperature. The number of
sites is half the number of particles since we are working with a
bilayer. This value stabilizes at area expansions near 4{\%} for
systems larger than $10^{4}$ sites. The dashed line indicates the
size of the network used in the rest of the simulations.}
\end{figure}

\begin{figure}
\resizebox{3.25in}{2.79in}{\includegraphics{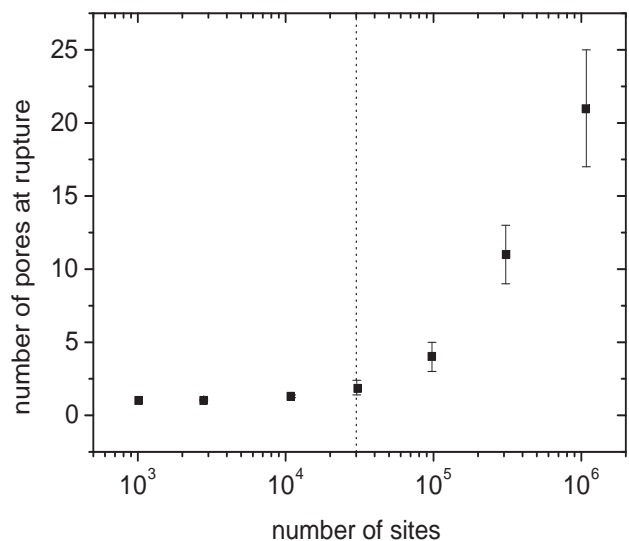}}
\caption{The average number of stable pores in the system when it
ruptures at room temperature. The relation is linear (exponential
on this log scale). The dashed line indicates the size of the
network used in the rest of the simulations.}
\end{figure}
\begin{figure}
\resizebox{3.25in}{2.63in}{\includegraphics{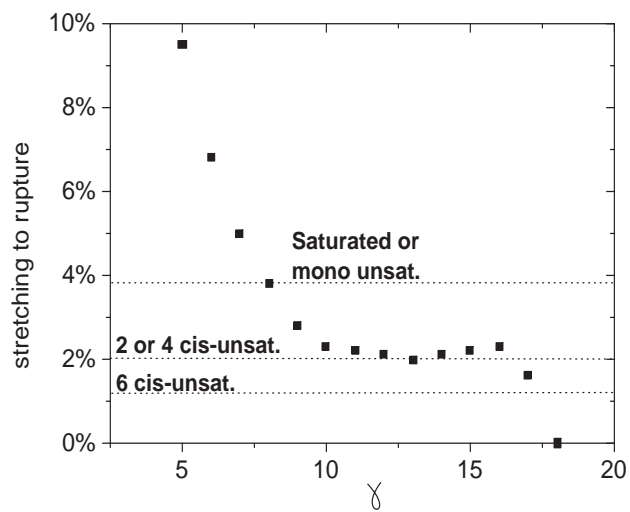}} \caption{
The degree of stretching for rupture at room temperature for
different tail rigidities for membranes with compressibility of
243 mN/m and membrane core thickness of 3,0 nm (see table 1). The
dashed lines are the known area expansions at rupture for
membranes with different degrees of unsaturation. }
\end{figure}
\begin{figure}
\resizebox{3.25in}{2.65in}{\includegraphics{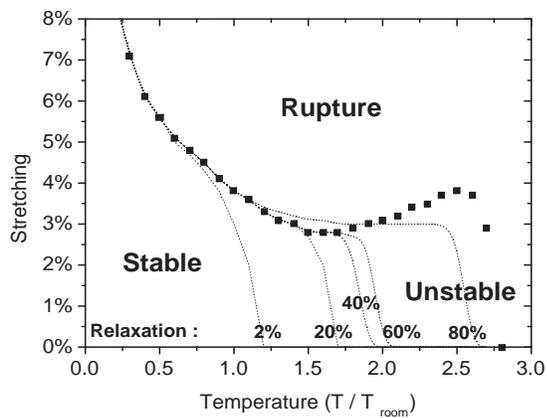}} \caption{
Phase diagram predicted by the model for a ''typical'' bilayer (se
Table I) showing the degree of relaxation in the membrane as a
function of stretching and temperature. The full square dots
indicate the rupture first-order transition line. The dashed lines
are the curves of constant relative relaxation. }
\end{figure}

\begin{figure}
\resizebox{3.25in}{2.65in}{\includegraphics{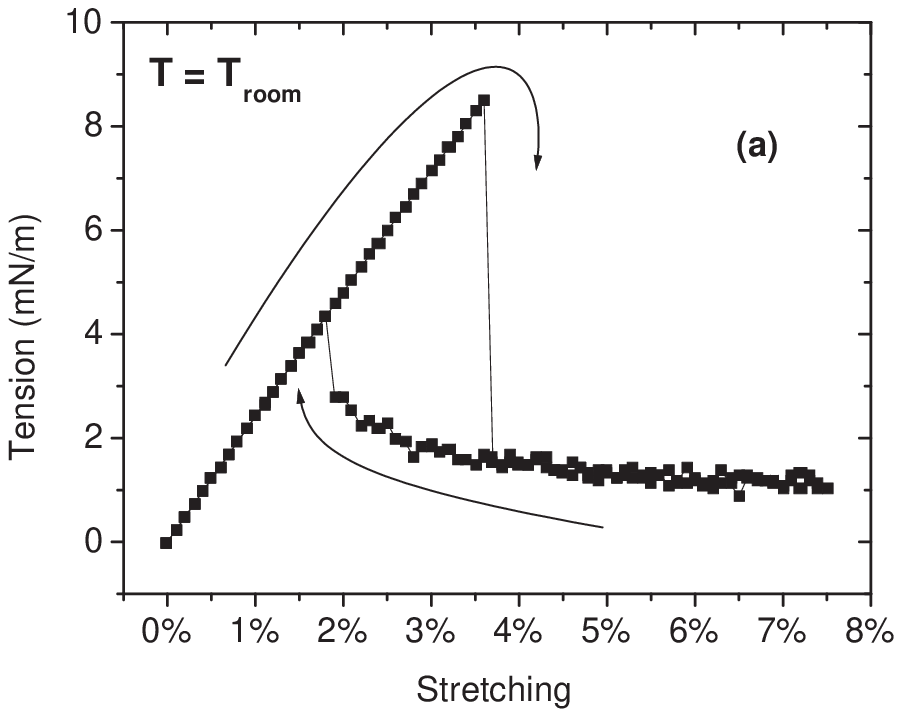}}
\resizebox{3.25in}{2.65in}{\includegraphics{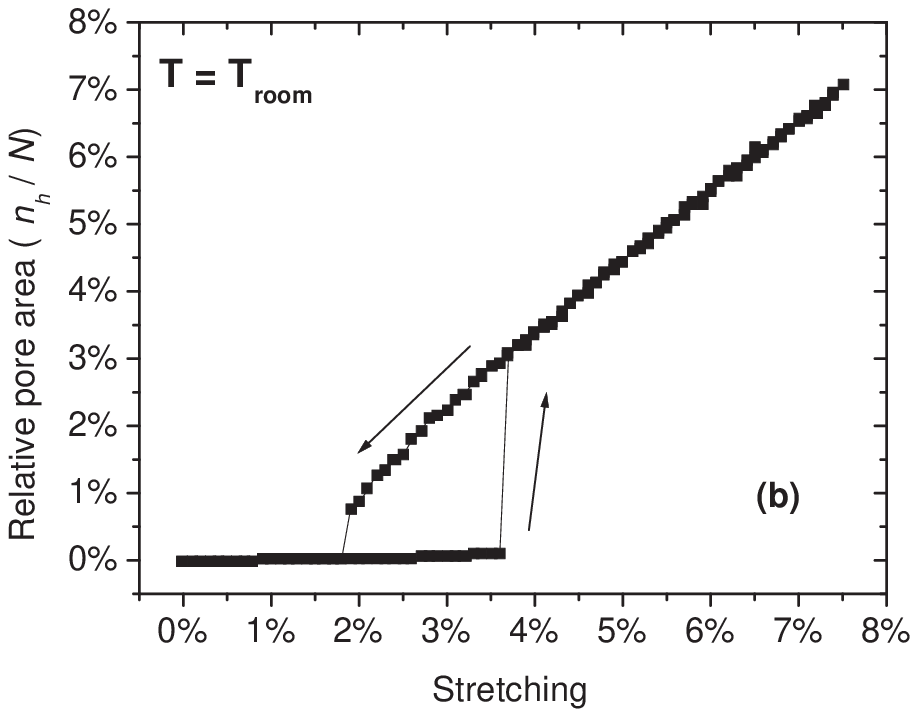}}
\resizebox{3.25in}{2.65in}{\includegraphics{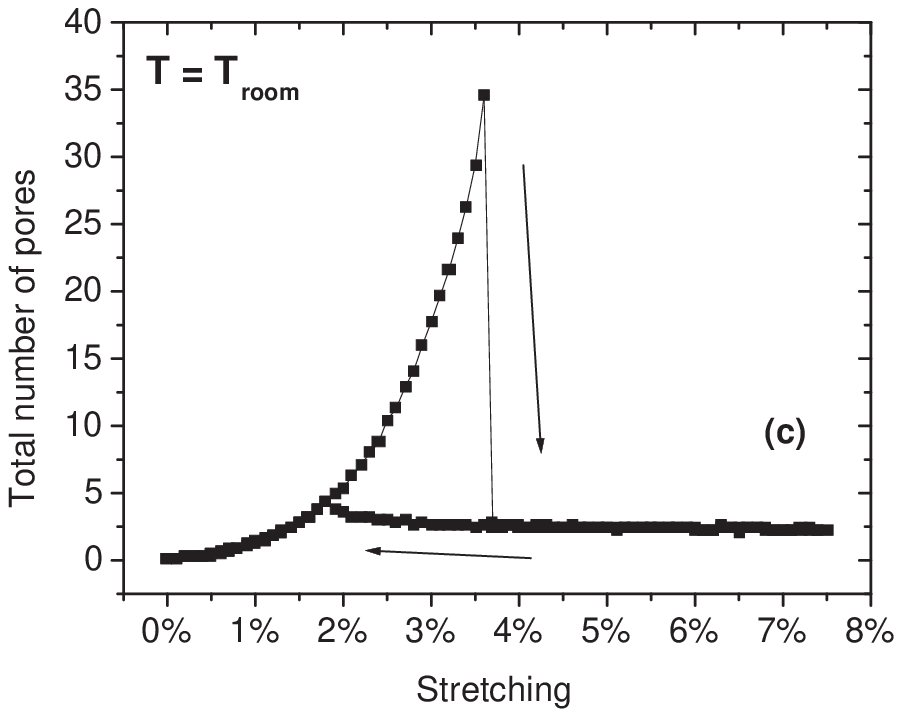}}
\caption{Hysteresis of different quantities upon stretching at $T
= T_{room}$: (a) the tension $\tau$; (b) the relative area
occupied by pores which is equal to the number of pore sites
divided by the total number of sites; (c) the total number of
pores. Room temperature systems behave similarly to high
temperature systems at low tensions by having a considerable
quantity of protopores present in the membrane. The rupture
scenario is however similar to that of lower temperature systems.
}
\end{figure}

\begin{figure}
\resizebox{3.25in}{2.65in}{\includegraphics{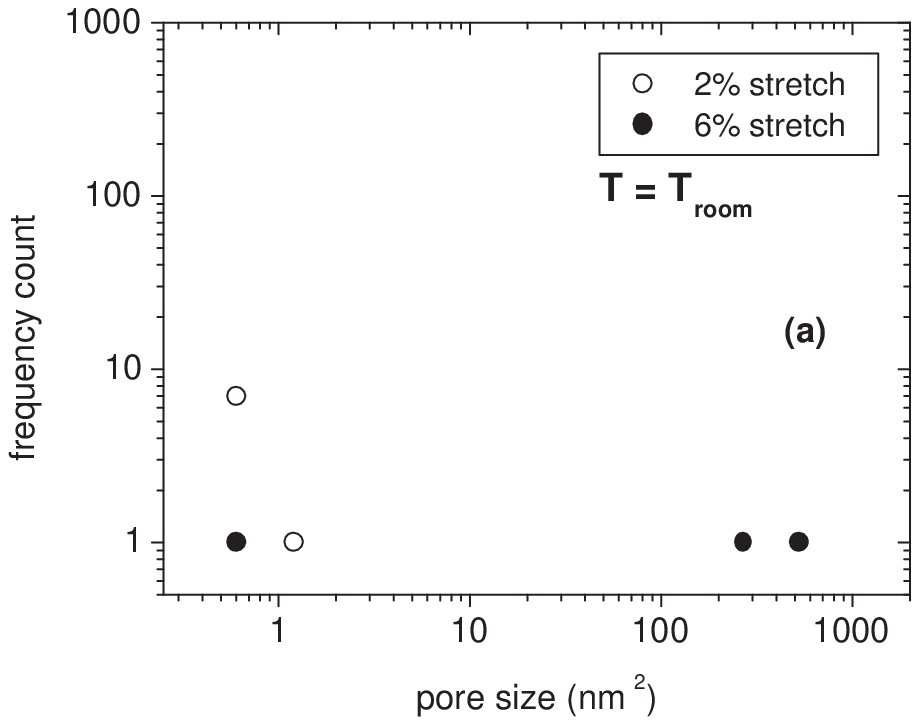}}
\resizebox{3.25in}{2.65in}{\includegraphics{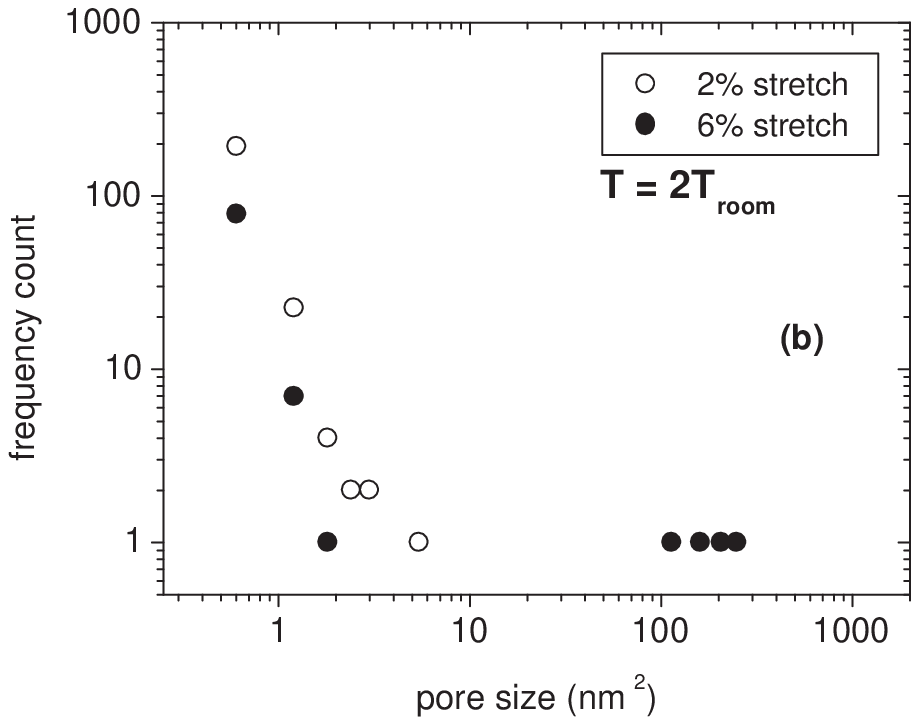}}
\resizebox{3.25in}{2.65in}{\includegraphics{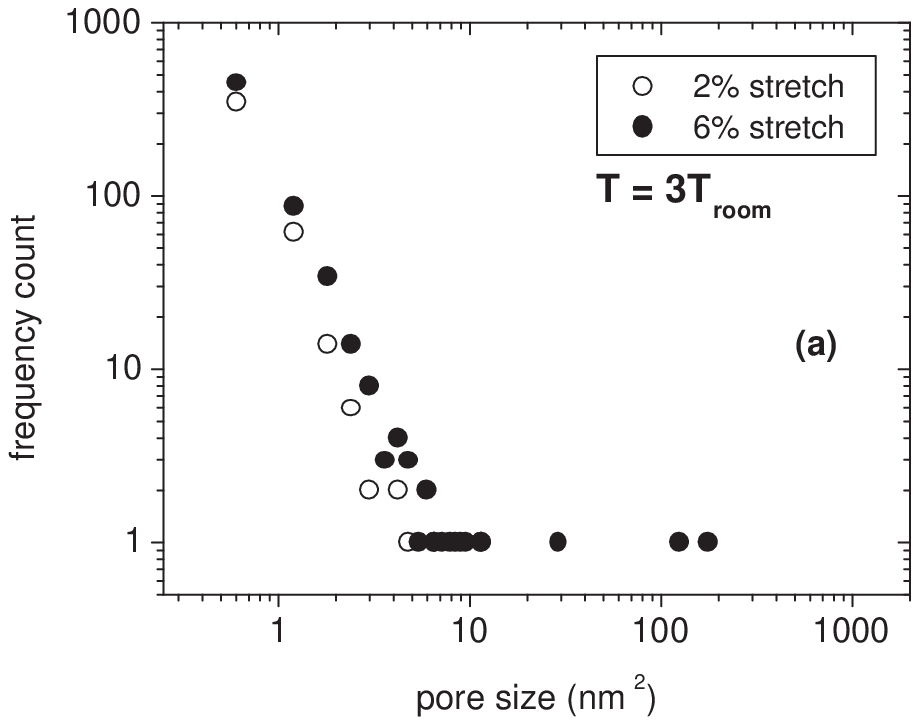}} \caption{
Distribution of pores at three temperatures (a) $T=T_{room}$, (b)
$T=2T_{room}$, and (c) $T=3T_{room}$, below and above the rupture
point in a lattice of 30301 sites (each site $\approx 0.6$
nm$^2$). With increasing temperature, the number of small pores
increases and a larger number remain in the system after rupture.
At $T=3T_{room}$, above the instability point of $2.8T_{room}$,
there is little structural difference between a $2\%$ and $6\%$
stretched membrane except for the larger size of the pores. }
\label{fig-por}
\end{figure}

\begin{figure}
\resizebox{3.25in}{2.56in}{\includegraphics{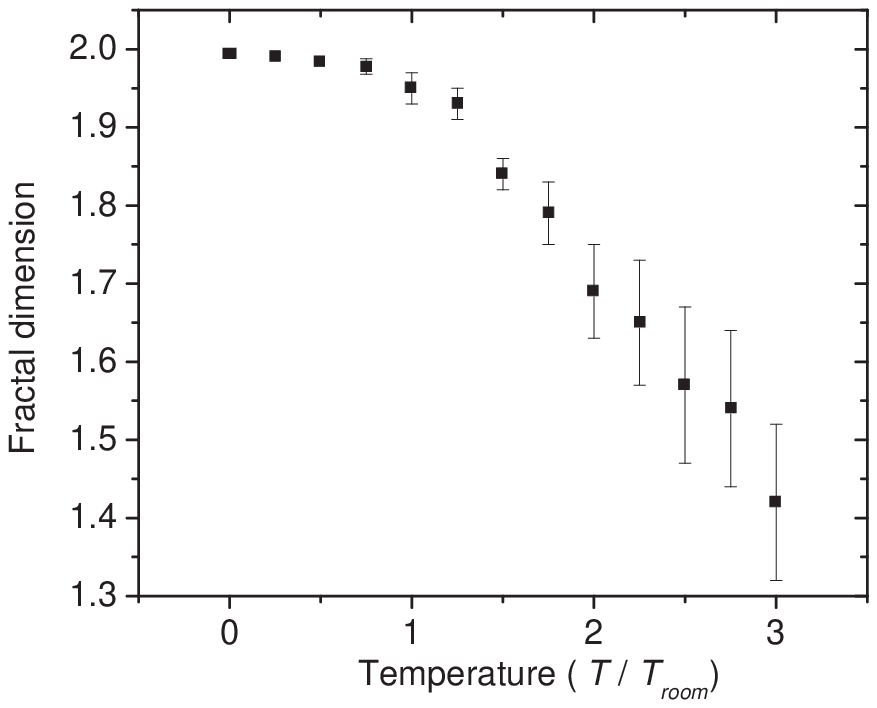}} \caption{
The Fractal dimension of a pore in a stretched membrane. This
quantity is a measure of the regularity of the shape of the pore
and its edge. As the temperature is increased, the pore becomes
irregular and spreads over a larger membrane surface. }
\end{figure}


\begin{thebibliography} {99}
\bibitem{note}The two limiting values of $L^{ah}$ are
a monovacancy $L^{ah}=1$ per site, for a straight segment of pore,
$L^{ah}=2$. For small pores with all its $n_i$ sites neighboring a
lipid site, $L^{ah}=6+2n_i$ and the smallest perimeter is equal to
$6+2(n_i-1)$ sites.
\end{thebibliography}
\end{document}